\documentclass{aastex63}
\usepackage{graphicx}
\usepackage{amsmath}
\usepackage{hyperref}
\usepackage{float}

\def\iso#1#2{\mbox{${}^{#2}{\rm #1}$}}
\def\fe6#1{\iso{Fe}{6#1}}
\def\al2#1{\iso{Al}{2#1}}
\def\mn5#1{\iso{Mn}{5#1}}
\def\be1#1{\iso{Be}{1#1}}

\def\msol{\mbox{$M_\odot$}}

\def\avg#1{\langle #1 \rangle}

\def\beq{\begin{equation}}
\def\eeq{\end{equation}}
\def\beqar{\begin{eqnarray}}
\def\eeqar{\end{eqnarray}}
\def\pfrac#1#2{\left( \frac{#1}{#2} \right)}
  
\def\mej{\mbox{$M_{\mathrm{ej,60}}$}} 
\def\df{\mbox{$f_{60}$}} 
\def\fobs{\mbox{$\mathcal{F}_{\mathrm{obs}}$}} 
\def\tarr{\mbox{$t_{\mathrm{arr}}$}} 
\def\trav{\mbox{$t_{\mathrm{trav}}$}} 
\def\up{\mbox{$U_{60}$}} 
\def\astropar{\mbox{$f_{60} \times M_{\mathrm{ej,60}}$}}

\begin{document}

\title{Distances to Recent Near-Earth Supernovae From Geological and Lunar \fe60}

\author[0000-0002-3876-2057]{Adrienne F. Ertel}
\affil{Department of Astronomy, University of Illinois, Urbana, IL 61801, USA}
\affil{Illinois Center for the Advanced Study of the Universe, University of Illinois, Urbana IL 61801}
\author[0000-0002-4188-7141]{Brian D. Fields}
\affil{Department of Astronomy, University of Illinois, Urbana, IL 61801, USA}
\affil{Department of Physics, University of Illinois, Urbana, IL 61801}
\affil{Illinois Center for the Advanced Study of the Universe, University of Illinois, Urbana IL 61801}

\begin{abstract}
Near-Earth supernova blasts which engulf the solar system have left traces of their ejecta in the geological and lunar records.  There is now a wealth of data on live radioactive \fe60 pointing to a supernova at 3 Myr ago, as well as the recent discovery of an event at 7 Myr ago.  We use the available measurements to evaluate the distances to these events.  For the better analyzed supernova at 3 Myr, samples include deep-sea sediments, ferromanganese crusts, and lunar regolith; we explore the consistency among and across these measurements, which depends sensitively on the uptake of iron in the samples as well as possible anisotropies in the \fe60 fallout. There is also significant uncertainty in the astronomical parameters needed for these calculations.  We take the opportunity to perform a parameter study on the effects that the ejected \fe60 mass from a core-collapse supernova and the fraction of dust that survives the remnant have on the resulting distance.  We find that with an ejected \fe60 mass of $3\times10^{-5}$ \msol\ and a dust fraction of 10\%, the distance range for the supernova 3 Myr ago is $D \sim 20 - 140$ pc, with the most likely range between $50 - 65$ pc.  Using the same astrophysical parameters, the distance for the supernova at 7 Myr ago is $D \sim 110$ pc.  We close with a brief discussion of geological and astronomical measurements that can improve these results.
 
\end{abstract}


\section{Introduction} \label{sec:intro}

In the last few decades, two global \fe60 ($t_{1/2}$ = 2.62 Myr\footnote{Half-life measurement: \citealp{rugel2009,wallner2015a,ostdiek2017}}) signals corresponding to near-Earth supernovae have been discovered in ferromanganese (FeMn) crusts and deep-sea sediments at 3 and 7 million years ago (Mya) \citep{knie1999, knie2004, fitoussi2008, ludwig2016,  wallner2016, wallner2021}.  An excess of \fe60 above the natural background has also been discovered in lunar regolith \citep{fimiani2016}.  The progenitors of these signals are most likely either core-collapse (CCSN) or electron-capture (ECSN) supernovae, as other producers of \fe60, such as thermonuclear supernovae and kilonovae, do not produce sufficient \fe60 mass to be within a plausible distance of Earth \citep{fry2015}.  Although not entirely ruled out, super-asymptotic-giant-branch (SAGB) stars are not considered in this paper, as their slow winds last a relatively short duration and do not match the observed \fe60 fallout timescale of $\gtrsim$ 1 Myr \citep{ertel2023}.  The two near-Earth supernovae conveniently fall into separate geologic epochs, and therefore we will refer to them as the Pliocene Supernova (SN Plio, 3 Mya) and the Miocene Supernova (SN Mio, 7 Mya). 
For recent reviews on near-Earth supernovae, see \citet{korschinek2023}, \citet{wallner2023}, \citet{fields2023}.

\citet{fry2015} used the available data from \citet{knie2004} and \citet{fitoussi2008} to put bounds on the distance from Earth to SN Plio given the observed \fe60 fluence.  Using the supernova \fe60 yields available at the time, they found a distance of $D \sim 60-130$ pc for CCSN and ECSN. We seek to expand on those calculations, given the plethora of new \fe60 data for SN Plio presented in \citet{ludwig2016}, \citet{fimiani2016}, \citet{wallner2016}, and \citet{wallner2021}.  In addition to the Earth-based data, the distance to the supernovae depends on three astronomical parameters: the ejected \fe60 mass from the progenitor, the time it takes for the dust to travel to Earth, and the fraction of the dust which survives the journey.  The time is thus ripe to investigate the impact those parameters have on the supernova distance.

The structure for the paper is as follows.  In Section~\ref{sec:math}, we lay out the relevant variables and their theory and data sources.  In Section~\ref{sec:data}, we examine the different \fe60 samples and their possible constraints.  In Section~\ref{sec:models}, we examine the three main astronomical parameters, their bounds, and the implications of their range on the supernova distance.  We then systematically map out the uncertainties in those parameters in Section~\ref{sec:results} and rule out model combinations.  In Section~\ref{sec:discussion}, we discuss other methods for calculating the supernova distance.

\section{Formalism}\label{sec:math}

The nucleosynthesis products from a supernova --- including radioisotopes such as \fe60 --- are ejected in the explosion and eventually spread throughout the remnant.  The time-integrated flux, or fluence, thus allows us to connect the observed parameters of the \fe60 signal on Earth with the astronomical parameters of the supernova remnant.  
In reality, the distribution of \fe60 in the remnant will be aniostropic, and the time history of its flux on Earth will be complex.  Because the supernova blast cannot compress the solar wind to 1 au without being within the kill (mass extinction limit) distance \citep{fields2008,miller2022}, the terrestrial signal can arise only from the ejecta arriving in the form of dust grains \citep{benitez2002,athanassiadou2011,fry2015,fry2016}.  We have argued that supernova dust decouples from the blast plasma, and that its magnetically-dominated propagation and evolution naturally lead to the observed $> 1 \ \rm Myr $ timescale for \fe60 deposition \citep{fry2020,ertel2023}.  The Earth's motion relative to the blast will also affect the \fe60 flux onto Earth \citep{chaikin2022}.

The \fe60 flux $\Phi_{60}(t)$ accumulates in natural archives over time.  This signal integrates to give the \fe60 fluence ${\cal F} = \int \Phi_{60}(t) \ dt$, which will be the central observable in our analysis.  Our goal in this paper, as with earlier distance studies \citep{fields1999,fry2015}, is not to capture all of this complexity, but to find a characteristic distance based on a simplified picture of a spherical blast engulfing an stationary Earth.

For a spherical supernova blast, we can generalize the relationship between the observed fluence of a radioisotope $i$ and the supernova properties as
\begin{equation} \label{eq:fluence}
    \mathcal{F}_{\mathrm{obs},i} = \dfrac{1}{4}\dfrac{M_{\mathrm{ej},i}}{4 \pi D^2 A_i \, m_u} \, U_i f_i \, \mathrm{exp}\left[\dfrac{-(t_{\mathrm{arrive}} + t_{\mathrm{trav}})}{\tau_i}\right] ,
\end{equation}
\noindent where $A_i$ is the mass number, $m_u$ is the atomic mass unit, and $\tau_i$ is the lifetime of the isotope.
The leading factor of 1/4 is the ratio of Earth's cross sectional area to surface area.
The two Earth-based parameters are the arrival time $t_{\rm arrive}$, which is the time of the first non-zero signal point, and the uptake fraction $U_i$, which quantifies the difference between the amount of the isotope that arrives at Earth and what is detected (see Section~\ref{sec:data}).  The four astronomical parameters are the ejected mass of the isotope $M_{\mathrm{ej},i}$, the fraction of the isotope that is in the form of dust $f_i$, the distance to the supernova $D$, and the travel time $t_{\rm trav}$ between the supernova and Earth.  Note that Eq.~(\ref{eq:fluence}) assumes a uniform fallout of the \fe60 onto Earth \citep[but see][]{fry2016}. 

Equation~\ref{eq:fluence} gives an inverse square law for the radioisotope fluence as a function of distance, similar to the inverse square relation for photon flux.  Setting aside the travel time's dependence on the distance, we can then solve Eq.~(\ref{eq:fluence}) as:
\begin{equation} \label{eq:dist}
    D = \dfrac{1}{2} \pfrac{\df \, \mej}{4 \pi A_{60} \, m_u}^{\frac{1}{2}} \, \pfrac{\up}{\fobs}^{\frac{1}{2}} \, \mathrm{exp}\left[\dfrac{-(t_{\mathrm{arrive}} + t_{\mathrm{trav}})}{2\,\tau_{60}}\right]. 
\end{equation}
Equation~(\ref{eq:dist}) is the main equation of interest in this work; therefore we have substituted the generic isotope $i$ for \fe60, as this is the isotope measured on Earth.  In the interest of brevity, $\mathcal{F}_{\rm obs,60}$ will be referred to as \fobs.  Note that \fobs\ is the fluence of \fe60 into the material (deep-sea sediment, FeMn crust, lunar regolith) and not the \fe60 fluence at Earth's orbit or in the interstellar medium (ISM) --- these latter two are a geometric factor of 4 different due to surface area and include corrective values such as the uptake factor. 

Equation~(\ref{eq:dist}) shows that distance scales as $D \propto (\df \, {\cal N}_{60} / \fobs)^{1/2}$.  We see the fluence scaling $\fobs^{-1/2}$ and additional dimensionless factors counting the number ${\cal N}_{60} = \mej/A_{60} \, m_u$ of \fe60 atoms and correcting for the dust fraction $\df$.  Moreover, the analogy to photon flux is very close: this {\em radioactivity distance} is formally identical to
a luminosity distance, with (uptake corrected) \fe60 fluence playing the role of photon flux, and the product $\df \, \mej$ of dust fraction and yield playing the role of luminosity.

The error on the radioactivity distance depends on both data-driven and astrophysical values.  Because an objective of this work is to examine the effects that different ejected masses, dust fractions, and travel times have on the supernova distance, the errors associated with those values are not included in our calculations.  Therefore the quoted distance error will only be the result of the data-driven values of observed fluence and uptake factor.\footnote{The arrival time also has an associated error, however it is negligible in this calculation and therefore not included.}  

All observed fluences have well-defined statistical errors.  However most of the uptake factors are quoted as an approximation or assumed to be 100\% --- due to the lack of clarity and the large influence these errors have on the resulting distance, for the purpose of this paper we will not be including the uptake error explicitly into our calculations.  Rather, we will illustrate the effect of this systematic error by displaying results for a wide range of uptakes corresponding to values quoted in the literature. \textit{The errors quoted on all of our distance calculations therefore solely reflect the reported statistical error on the fluence}.

We can calculate the error on the radioactivity distance that arises due to uncertainties in the fluence.  This is simply
\begin{equation} \label{eq:derr}
    \sigma_D = \dfrac{1}{2}\, D\, \pfrac{\sigma_\mathcal{F}}{\fobs}, 
\end{equation}
\noindent where $\sigma_{\mathcal{F}}$ is the error on the observed fluence.  

It is important to note that the radioactivity distance scales as $D \propto \sqrt{\df \, \mej}$, so that the key astrophysics input or figure of merit is the product $\df \, \mej$ of \fe60 yield and dust fraction.  This represents physically the effective yield of \fe60 in a form able to reach the Earth.  The resulting radioactivity distance is therefore most affected by the allowed range of \mej\ and \df.  To that effect, this paper presents the quantity \astropar\ [\msol] as a means of approximating the maximum and minimum astronomical parameters that can be used to find a supernova distance's from Earth. 

\section{Data and Benchmark Results}\label{sec:data}

The data used in this analysis are from the work of \citet[][hereafter K04]{knie2004}, \citet[][F08]{fitoussi2008}, \citet[][L16]{ludwig2016}, \citet[][W16]{wallner2016}, \citet[][F16]{fimiani2016}, and \citet[][W21]{wallner2021}. The \fe60 signal has been found in a number of different materials on Earth, including deep-sea sediment cores and ferromanganese (FeMn) crusts, as well as in the lunar regolith.  

FeMn crusts are slow-growing, iron and manganese-rich layers which build up on exposed rock surfaces in the ocean at a rate of $\sim$ few mm/Myr.  Ferromanganese nodules have a similar growth rate and are found as individual objects on the sea floor. These crusts grow by extracting iron and manganese from the surrounding seawater, and thus they have an associated uptake factor, which accounts for how much of the available iron in the seawater they absorb.  The uptake factor varies considerably with each crust, on the order of $1-30$\% (see Tab.~\ref{tab:data}), and must be calculated for each sample.  In contrast, deep-sea sediments grow much faster rate of $\sim$ few mm/kyr.  Unlike FeMn crusts, they are assumed to have a 100\% uptake factor as they sample what is deposited on the ocean floor.

\begin{table}[t]
    \centering
    \caption{Data-driven values for calculating the distance to SN Plio.}
    \hspace{-25mm}
    \begin{tabular}{lc|ccc|c}
        \hline \hline 
         Paper &  & $\fobs\ [10^6 \, \rm atoms/cm^2]$ & \tarr\ [Myr] & \up & D [pc] \\
         \hline
         \citet{knie2004} crust\footnote{\fobs\ and \tarr\ are from the \citet{wallner2016} Tab.~3 (corrected for \fe60 and \be10 half-life changes since the publication of K04),  uptake is the original quoted in \citet{knie2004} (see Subsection~\ref{ssec:knie}).} & K04 & $1.5\pm0.4$ & 2.61 & $\sim 0.006$ & $22\pm3$ \\
         \citet{knie2004} crust &  & $1.5\pm0.4$ & 2.61 & $\sim 0.24$\footnote{Uptake is from discussion in \citet{fimiani2016} (see Subsection~\ref{ssec:knie}).} & $140\pm19$ \\
         \citet{knie2004} crust &  & $1.5\pm0.4$ & 2.61 & $\sim 0.04$\footnote{Uptake is calculated from the \citet{wallner2016} sediment (see Subsection~\ref{ssec:knie}).} & $57\pm8$ \\
         \hline
         \citet{fitoussi2008} sediment A & F08 & $30.0\pm14.5$\footnote{Fluence values are not specifically quoted in the original paper; these have been calculated using the area under the curve for Fig.~4A and 4B in \citet{fitoussi2008} (see Subsection~\ref{ssec:fitoussi}).} & 2.87 & 1.0 & $64\pm15$ \\
         \citet{fitoussi2008} sediment B &  & $58.0\pm39.0^{\rm d}$ & 3.08 & 1.0 & $46\pm15$ \\
         \hline
         \citet{ludwig2016} sediment & L16 & $0.56\pm0.18$ & 3.02\footnote{L16 states that the signal starts at 2.7 $\pm$ 0.1 Mya, however first not-zero binned point in their Fig.~2B is 3.02 Myr; we use the 3.02 Myr start here.} & 1.0 & $470\pm75$ \\
         \hline
         \citet{wallner2016} sediment & W16 & $35.4\pm2.6$ & 3.18 & 1.0 & $58.9\pm2.2$ \\
         \citet{wallner2016} crust 1 &  &  $5.9\pm0.8$ & 4.35 & $0.17\pm0.3$ & $59\pm4$ \\
         \citet{wallner2016} crust 2 &  & $2.2\pm0.2$ & 3.1 & $\sim 0.07$ & $62\pm3$ \\
         \citet{wallner2016} nodules &  & $1.4\pm0.5$ & 3.3 & $0.02-0.04$ & $51\pm9 $ \\
         \citet{wallner2021} crust 3 & W21 & $6.10\pm0.31$ & 4.2 & $0.17\pm0.3$ & $58.3\pm1.5$ \\
         \hline
         \citet{fimiani2016} lunar & F16 & $10-60$ & 2.6 & 1.0 & $45 - 110$ \\
         \hline \hline
    \end{tabular}
    \label{tab:data}
    \vspace{2mm}
    \parbox{0.9\textwidth}{\fobs\ is the fluence into the material (sediment, FeMn crust, FeMn nodule, or lunar regolith); \tarr\ is the arrival time of the dust on Earth, when the \fe60 signal starts; \up\ is the uptake percentage of \fe60 into the material, with sediments and lunar regolith assumed to have a 100\% uptake.  The example of a possible distance $D$ to SN Plio quoted here is calculated assuming $\mej \times \df = 3\times10^{-6}$ \msol, with $\mej = 3 \times 10^{-5}$ \msol\  and \df\ = 0.1.}
\end{table}

\begin{figure}[htb]
    \centering
        \includegraphics[width=16cm]{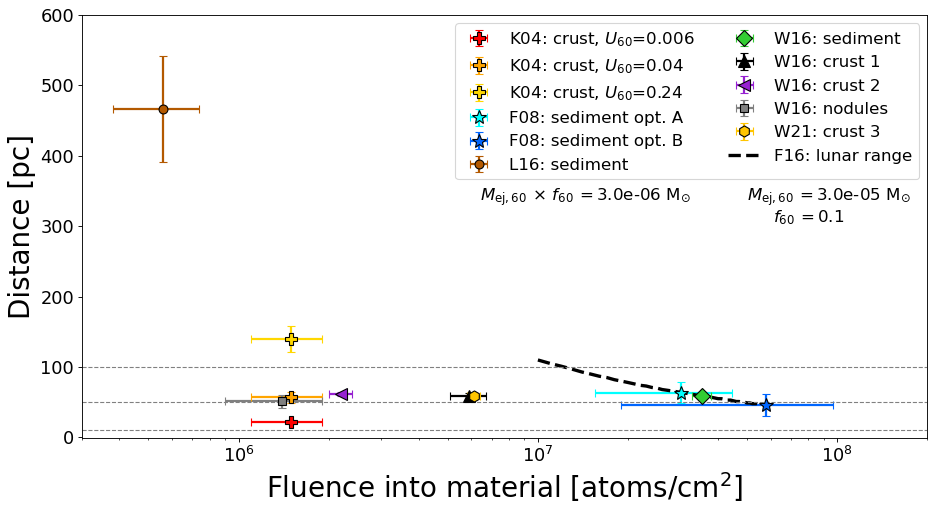} 
        \includegraphics[width=16cm]{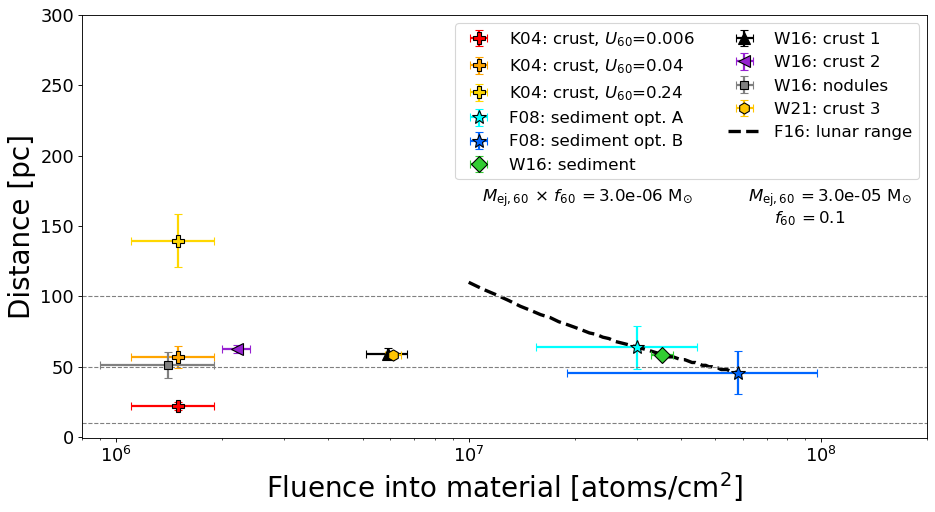} 
  \caption{\textit{Distance to supernova vs fluence into material for SN Plio (3 Mya)}.  The assumed \astropar\ is printed under the legend. Horizontal dotted lines show distances of 100, 50, and  10 pc (the kill distance).  \textbf{Top:} All of the \fe60 fluence data, including L16.  K04 is shown with three possible uptake factors (see Subsection~\ref{ssec:knie}), demonstrating the effect of the uptake on the distance.  F08 opt.~A and B are two different binnings for the same sediment data, not two sediments (see Subsection~\ref{ssec:fitoussi}).  Note that all of the W16 and W21 data, as well as K04 (\up=0.4, orange cross), are correlated to the W16 sediment data and therefore show approximately the same distance to SN.  \textbf{Bottom:} A zoom in on the distances without the L16 point. For paper citations and abbreviations, as well as calculation details, see Tab.~\ref{tab:data}.}
  \label{fig:dist}
\end{figure}

Table~\ref{tab:data} summarizes the observed fluences (\fobs), the \fe60 arrival times (on Earth and the Moon), and the uptake percentage of \fe60 into the material for all of the \fe60 detections considered in this work.\footnote{A low-level \fe60 infall over the last 30 kyr has been measured by \citet{koll2019} and \citet{wallner2020}; however, we do not consider it as part of the same astrophysical delivery mechanism that created the \fe60 peaks considered in this work and therefore this infall is not included.}  
We see that the arrival times are for the most part quite consistent, even across crust and sediment measurements.

Table \ref{tab:data} also provides an example of a distance to SN Plio.  These results all use the quoted fluence and uptake, and assume $\astropar = 3\times 10^{-6} \msol$. In the next sections, we will address in detail the correlations between these results and the wide variety of distances they give.  The range of distances is much larger than the quoted statistical errors, confirming that systematic errors --- most notably the uptake --- dominate the distance uncertainties.

Figure~\ref{fig:dist} then plots the distance vs fluence for the published \fe60 data relating to SN Plio.  The distance is calculated as shown in Tab.~\ref{tab:data} and the fluence refers to the fluence into the material on Earth.
Error bars on the fluence are as quoted in the original papers; error bars on the distance trace the fluence error effects.\footnote{As can be seen in Tab.~\ref{tab:data}, only the W16 crust 1 and W21 crust 3 have precise errors on the uptake (the sediment and lunar uptakes are assumed to be 100\% and thus do not have an associated error).  Without more precise values for the other FeMn crust uptakes and in the interest of consistency between datasets, we ignore all uptake errors here.}  The top plot of Fig.~\ref{fig:dist} shows all of the data, while the bottom plot neglects the outlier L16 sediment data (discussed in Subsection~\ref{ssec:ludwig}). 

Figure~\ref{fig:dist} represents a consistency check among the \fe60  measurements.  The reported fluence and uptake are used to infer the interstellar fluence arriving at Earth, and this in turn leads to the distances plotted.  As seen in eq.~(\ref{eq:dist}), all results scale with the adopted yield and dust faction as $D \propto (\df \mej)^{1/2}$.  Because this factor is common to all points shown in the plot, the entire pattern can shift up or down systematically for different choices of this parameter combination.  But crucially, whether the distances we infer are consistent or discrepant does not depend on these parameter choices. 

We will review the agreement among data sets in detail below, but the main results are clear from a glance at Figure \ref{fig:dist}. We see that most results span 50 to 150 pc, which are shown in a zoom in the bottom panel.  There is a group of data clustered together in distance, from around 40 to 70 pc, which shows a non-trivial consistency --- though we will see that most of the points are correlated.  Note that the K04 crust results are shown for different uptake values, making it clear that this choice can lead to consistency (if $U_{60} \sim 0.04$ for this crust) or discrepancy (if  $U_{60}$ takes a substantially different value).
On the other hand, the top panel shows that the L16 results lead to distances that are far from the others.  We discuss this in detail below.

Horizontal lines on Fig.~\ref{fig:dist} indicate key astrophysical distances.  The lowest line at 10 pc is an estimate of the typical SN kill distance, inside of which substantial damage to the biosphere is expected \citet[e.g.,][]{gehrels2003,brunton2023}.
No points lie below this range, consistent with the lack of widespread anomalous biological extinctions in the past 3 Myr.  
The other two lines show the position of nearby star clusters that have been proposed to host SN Plio: the $\sim 50$ pc location of the Tucana-Horologium association, and the $\sim 100$ pc distance to the Scorpius-Centaurus association.  We see that the clustered data points are consistent with the Tuc-Hor distance, though a somewhat larger \astropar\ would favor Sco-Cen.
Finally, we note that the maximum size of a supernova remnant is can be estimated from the ``fadeaway distance'' \citep{draine2011} when the blast wave becomes a sound wave, which depends on the density of the ambient medium but is $\sim 100-150 \ \rm pc$.  We see that all of the points are inside this distance, as would be expected for a SN origin of \fe60 --- except for L16.  Thus, aside from L16, the \fe60 data is consistent with astrophyiscal expectations, which represents a non-trivial test, because astrophyical distances are not built into the \fe60 measurements (in contrast to the $\sim \ \rm Myr$ timescale that is pre-ordained by the choice of \fe60).  

We now examine the datasets and results in Fig.~\ref{fig:dist} in more detail.
There are three possible uptake factors to use for the K04 data (see Subsection~\ref{ssec:knie} and Tab.~\ref{tab:data}) and all three have been included as separate date points to demonstrate the effect of the uptake factor on the supernova distance.  The F08 data have two points to represent the two options presented in Tab.~4 of \citet{fitoussi2008}; these are the same sediment sample fluence calculated two different ways, not independent measurements (see Subsection~\ref{ssec:fitoussi}).  The uptake factors for the W16 and W21 crusts and nodules are calculated based on the assumption that the W16 sediment collected 100\% of the \fe60 fluence, and so all of the W16 and W21 data are correlated and trace approximately the same distance (see Subsection~\ref{ssec:wallner}).  The K04 point with \up = 0.04 (orange cross in Fig.~\ref{fig:dist}) is similarly calculated and therefore also correlated with the W16 and W21 data.  

The F16 lunar data are plotted as a dashed line showing the full possible range quoted in \citet{fimiani2016}.  As described in Subsection~\ref{ssec:lunar}, the time window for this fluence covers the last 8 Myr, during which there have been two near-Earth supernovae at 3 Mya and 7 Mya.  However, the 3 Mya supernova contributes 90\% of the observed fluence (as calculated in Subsection~\ref{ssec:lpw21}) and therefore the dashed line more or less traces the available distance and fluence range for SN Plio.

It is of note that the W16 sediment fluence and the two possible F08 fluences fall on the lunar fluence line, with the significantly more precise W16 sediment in exact agreement.  The full implications of this alignment are analyzed in Subsection~\ref{ssec:lunar}, but altogether it does lend credence to the idea that the deep-sea sediments are sampling 100\% of the \fe60 flux which falls on the Earth.

\subsection{Knie+ 2004 Data (K04)} \label{ssec:knie}

The K04 FeMn crust was the first measurement of the \fe60 signal from SN Plio that provided a time profile.  Since the paper was published, the \fe60 and \be10 half-lives have been updated: the \fe60 half-life has changed from 1.47 Myr \citep{kutschera1984} to 2.62 Myr \citep{rugel2009,wallner2015a,ostdiek2017}; while the \be10 half-life has changed from 1.5 Myr to 1.36 Myr \citep{nishiizumi2007} to 1.387 Myr \citep{korschinek2010}. \citet{wallner2016} updates K04's fluence for these half-life changes in their Tab.~3 and we use those numbers here.  It should be noted that the same FeMn crust was measured by F08, which confirmed the \fe60 signal results.

FeMn crusts do not absorb all of the available iron in the seawater they contact; thus, it is necessary to calculate an uptake efficiency factor, \up, for the crust. Unfortunately, this factor cannot be measured directly and must be inferred.  K04 cites \citet{bibron1974} as a means of calculating the uptake factor for the FeMn crust.  By using the known \mn53 extraterrestrial infall and comparing elemental ratios of Mn and Fe in seawater to the \mn53 found in the FeMn crust, K04 was able to estimate the Fe uptake factor.  As explained in F16 and confirmed by T. Faestermann and G. Korschinek (private communication), recent work with the \mn53 infall corrects \citet{bibron1974} by a factor of 40 smaller.  The factor of 40 decrease in \mn53 leads to a relative factor of 40 increase in the \fe60 to match the \fe60/\mn53 ratio detected in the crusts, and thus the \fe60 uptake for the FeMn crust in K04 changes from 0.6\% to 24\%.

An alternative method to calculate the uptake factor for the crust is to use a known \fe60 infall over the relevant period of time, such as the W16 sediment.\footnote{This method was not available for the original calculation, as \fe60 would not be measured in sediments until F08.}  By dividing the K04 incorporation by the W16 sediment incorporation in Tab.~3 of W16, we find an uptake factor of 4\%, consistent with the FeMn crust and nodule uptakes in W16 and W21, which use the same method.  This method correlates all of the distance measurements that are based off of the W16 sediment, leaving only the F08, L16, and F16 as independent measurements. 

To demonstrate the importance of the uptake factor in the distance calculation, the three options for the K04 uptake (\up = 0.6\%, \up = 4\%, \up = 24\%) are shown in Tab.~\ref{tab:data} and Fig.~\ref{fig:dist}.

\subsection{Fitoussi+ 2008 data (F08)}\label{ssec:fitoussi}

F08 measured \fe60 in both FeMn crusts and in deep-sea sediments.  They first repeated the \fe60 analysis on the same crust as used by K04, confirming the \fe60 peak; since that work recreates an existing measurement, we do not use those results in this paper.  F08 also pioneered the first \fe60 analysis on deep-sea sediment samples.  They found no significant signal unless they binned their data using a running-means average of either 0.4 or 0.8 Myr, as shown in their Fig.~4.  

The theory at the time was the that the \fe60 was in dust following the SN blast wave, which would take about 10 kyr to sweep over the solar system \citep{fry2015}.  The \fe60 timescale they were looking for (to match the fluence seen in K04) was actually spread over $\gtrsim$ 1 Myr \citep{ertel2023}, as would be shown in later work such as L16 and W16 --- thus greatly diluting the signal they expected to find.  Although the F08 sediment data cannot be used for a reliable time profile, we are able to include it in this work, as we are interested in measuring the fluence of their data and not the specific timing details.  

Using the two running-means averages shown in Fig.~4 of F08, we calculate the area under the curve and thereby the fluence by fitting a triangle to the upper plot (A) and two back-to-back triangles to the lower plot (B).  To find a fluence comparable to what is shown in other work, we:  1) updated the \be10 half-life from 3.6 Myr to 3.87 Myr \citep{korschinek2010} and changed the timescale accordingly; 2) subtracted the background of $2.3 \times 10^{-16}$ from the \fe60/Fe ratio; and 3) decay corrected the \fe60/Fe ratio using $t_{1/2} = 2.62$ Myr.  

From there, we were able to calculate the fluence for F08 via

\begin{equation}
\fobs = \int \frac{\fe60}{\rm Fe} \ \rho \, \dot{h} \, c_{\rm Fe} \ \, dt,
\end{equation}

\noindent where $\rho = 1.6$ g/cm$^3$ is the sediment density, $\dot{h} = 3000$ cm/Myr is the sedimentation rate, and $c_{\rm Fe} = 5.39 \times 10^{19}$ atoms/g is the iron concentration in the sediment, corresponding to a mass fraction of $0.5 \ \rm{wt\%}$.  Errors were pulled from the 1$\sigma$ lines around the running-averages in Fig.~4a and 4b of F08.  The results are shown in Tab.~\ref{tab:data}, labeled A and B to match the relevant plots in the original paper.  

\subsection{Ludwig+ 2016 data (L16)}\label{ssec:ludwig}

The L16 sediment data is notable in that the group set out to answer a different research question than the other \fe60 analysis:  instead of measuring the total \fe60 over a specific time range, their goal was to prove that the iron was not moving around in the sediment column due to chemical processes.  To achieve this, they focused on analyzing microfossils in the sediment and discarded iron material $\gtrsim 0.1 \mu$m, using the assumption that, since the \fe60 is vaporized on impact with the atmosphere, there should not be any \fe60 in larger sized grains.  However, the resulting \fobs\ is notably 1 - 2 orders of magnitude smaller than the other deep-sea sediments measured in F08 and W16, as well as the range for the lunar fluence in F16.  As can be seen in Tab.~\ref{tab:data}, the low \fobs\ in L16 puts SN Plio at an implausibly far distance from Earth, bringing into question whether dust from a supernova at that distance can even reach our solar system \citep{fry2015,fry2020}.  While it is possible to manipulate the values for the dust fraction and ejected mass used in the distance calculation to bring the L16 sediment within a reasonable distance of Earth (ie: to within at most 150 - 200 pc away, see Subsection~\ref{ssec:dustfrac} and~\ref{ssec:mej}), this unfortunately pushes the rest of the observed \fe60 fluences within the 10 pc ``kill distance'' and is therefore not realistic.

L16 compare their data to the sediment fluence found in W16 and attribute the difference to global atmospheric fallout variations.  However, as noted in \citet{fry2015} and \citet{ertel2023}, latitude variations would account for a factor of 5 difference at the most --- this is not enough offset the differences in the observed fluences.  Furthermore, the sediment samples from F08 and W16 see similar fluences despite significant location differences (off the coasts of Iceland and Australia, respectively).  It should also be noted that, while latitude fallout variation may account for some range in \fobs, the supernova should still be at minimum within reasonable travel distance to Earth.  Therefore, we must assume that some of the discarded sample from L16 contained \fe60.

The work of L16 conclusively demonstrated that the \fe60 is not moving within the sediment column after being deposited.  This is a major contribution to the field, considering that the observed \fe60 data is spread over an order of magnitude longer timescale than what is conventional for a supernova shock-wave and there are considerable implications for this effect to be astronomical in origin rather than geophysical \citep{ertel2023}.  However, due to the fact that the \fe60 fluence results in a $>400$ pc supernova distance, we will not be using the numbers from L16 in this study.

\subsection{Wallner+ 2016 and 2021 data (W16 and W21)} \label{ssec:wallner}

W16 measured the 3 Mya \fe60 signal in two FeMn crusts, two FeMn nodules, and four deep-sea sediments.  They greatly increased the known evidence of the signal and were also able to find indications of a second \fe60 signal 7 Mya.  W21 followed up these measurements in a separate FeMn crust and were able to verify the 7 Mya SN signal as well as provide an excellent time profile of both SN.\footnote{W21 notes that the observed time profile in the crust is wider than anticipated for SN Plio (which is profiled in the W16 sediments), indicating that factors such as crust porosity could affect these results.}

W16 and W21 calculated the uptake factors for their FeMn crusts and nodules by assuming that their sediment samples observed 100\% of the flux of \fe60 onto Earth.\footnote{\citet{koll2019} and \citet{wallner2020} both study the current low-level \fe60 infall, found in Antarctic snow and the top layers of the W16 sediments, respectively.  Both groups find approximately the same current flux of \fe60 --- considering that these are very different sample types in different environments and global locations, this could be used as strong evidence that the 100\% uptake assumption for deep-sea sediments is accurate, and that the W16 sediment data does accurately express the global \fe60 fallout for SN Plio.}  This connection between the different datasets means the resulting distance numbers are entirely correlated --- as seen in Tab.~\ref{tab:data} and Fig.~\ref{fig:dist}, the W16 and W21 data trace the same SN distance.  When using the K04 uptake as 4\% based off the W16 sediment (as in Fig.~\ref{fig:comp}), the K04 data is similarly correlated. 

It should be noted that the W16 and the W21 quote ``deposition rate'' and ``incorporation rate'', respectively, instead of a fluence.  This is the fluence into the material, which we use throughout this paper.  To connect it to the fluence into the solar system that is quoted later in W16 and W21, factors such as uptake and global surface area must be accounted for and corrected out of the equation.

W21 measured the \fe60 in an FeMn crust from 0 Mya to 10 Mya using sample slices of $\sim 400$ kyr.  In doing so, they created a detailed time profile showing two \fe60 peaks in the last 10 Myr, which can be attributed to two supernovae.  The peaks were measured in the same sample using the same analytical techniques --- thus if we compares their relative fluences, most of the geophysical 
complications and systematic errors drop out of the results.  Only issues such as fluctuations in the growth rate over millions of years and other large shifts in absorption into the crust over long periods of time will influence the results.  

\subsection{Fimiani+ 2016 Data (F16)} \label{ssec:lunar}

Unlike the Earth-based samples, the lunar samples analyzed by F16 are not affected by atmospheric, geologic, or biologic processes.  They also present an analysis that is fully independent from anything measured on Earth and which can be used to verify the many different techniques and sample types involved in analyzing the \fe60 signals.  

When using the lunar data, we must work with two effects caused by the lack of atmosphere:  cosmic ray nucleosynthesis and micrometeorite impacts.  The solar and galactic cosmic rays create a natural background of \fe60 in the lunar regolith and any \fe60 signal related to SN Plio will be shown in an excess of \fe60 above the standard background.  In addition, the micrometeorite impacts create a ``lunar gardening'' effect that churns the top regolith and makes time resolution under 10 Myr ambiguous \citep{fimiani2016,costello2018}.  In previous work, gardening was not an issue, as the excess \fe60 in the $\sim 8$ Myr 
sample was attributed to SN Plio; however, W21 has shown that there are actually two near-Earth supernovae in the last 10 Myr, at 3 Mya and 7 Mya.  Therefore, the excess \fe60 in the lunar regolith accounts for both supernovae, and in order to accurately calculate the distance to SN Plio using the lunar data, we first need to portion the excess \fe60 signal between the two supernovae.

\subsubsection{Data-driven portioning with the W21 results} \label{ssec:lpw21}

With the data from W21, we have an \fe60 signal that goes back 8 Myr and shows two distinct supernova peaks.  By taking the fluence ratio of these peaks, we are able to portion the lunar \fe60 signal into two separate supernovae.  W21 has the \fe60 fluence for SN Plio (3.1 Mya)  $\fobs = 6.10 \times 10^{6} \, \mathrm{atoms/cm}^2$ and for SN Mio (7.0 Mya)  $\fobs = 1.77 \times 10^{6} \, \mathrm{atoms/cm}^2$, both of which are decay corrected.\footnote{W21 quotes an ``incorporation rate'' which is proportional to the fluence; however, we are only interested in the ratio between these two values and thus the difference falls out.  Furthermore, since the two supernova peaks were measured in the same data slice from the same FeMn crust, the systematic and geo-related errors (such as uptake factor, various Earth processes that affect the signal, and any errors with absolute timing) cancel out.}  The F16 lunar data is also decay corrected, under the assumption that the excess \fe60 signal seen in the 8 Myr sample was actually deposited 2.6 Mya (to match the K04 FeMn crust signal).  We now know that two supernovae occurred within the last 8 Myr and therefore this decay correction needs to be fixed.  

To portion the lunar signal, the first step is to undo the decay correction on all three fluences using
\begin{equation}\label{eq:decaycor}
    \mathcal{F}_0 = \mathcal{F} \ 2^{-t_{\rm arr}/t_{1/2}} \,,
\end{equation}
\noindent where $\mathcal{F}_0$ is the ``dug up'' fluence, $\mathcal{F}$ is the decay corrected fluence, $t_{\rm arr}$ is the time that was used to decay correct the fluence (which in this case is the expected arrival time of the \fe60 signal), and $t_{1/2}$ is the half-life of the isotope (2.62 Myr for \fe60).  $t_{\rm arr}$ = 2.6 Myr for the lunar signal F16 and $t_{\rm arr}$ = 3.1 Mya and 7.0 Mya for the two W21 crust signals corresponding to SN Plio and SN Mio.  From there, we can calculate the respective ratio of the two supernovae fluences to each other, with
\begin{align}\label{eq:portion}
    \mathcal{P}_{\rm Plio} &= \dfrac{1}{\left(\mathcal{F}_{\rm Mio}/\mathcal{F}_{\rm Plio} \right) + 1} \\
    \mathcal{P}_{\rm Mio} &= 1-\mathcal{P}_{\rm Plio} = \frac{\mathcal{F}_{\rm Mio}}{\mathcal{F}_{\rm Plio}}\, \mathcal{P}_{\rm Plio}\ ,
\end{align}
\noindent where $\mathcal{P}_{\rm Plio}$ and $\mathcal{P}_{\rm Mio}$ are the percentages of the fluence from each supernova. We find that about 90\% of the excess lunar \fe60 signal should come from SN Plio (3.1 Mya), and 10\% should come from SN Mio (7.0 Mya).  We can then portion the ``dug up'' lunar fluence range of $1 - 6\times 10^6$ atoms/cm$^2$ and redo the decay correction, using $t_{\rm arr} = 3.1$ Myr for SN Plio and $t_{\rm arr} = 7.0$ Myr for SN Mio \citep{wallner2021}.  From there, we can calculate the distances to the two supernovae using Eq.~(\ref{eq:dist}).  It should be noted that this calculation assumes the same \mej, \df, and travel time for both SN; these values, possible ranges, and effects on the distance are examined in detail in Sections~\ref{sec:models} and~\ref{sec:results}.  

\begin{figure}[ht]
	\centering
  \includegraphics[width=17cm]{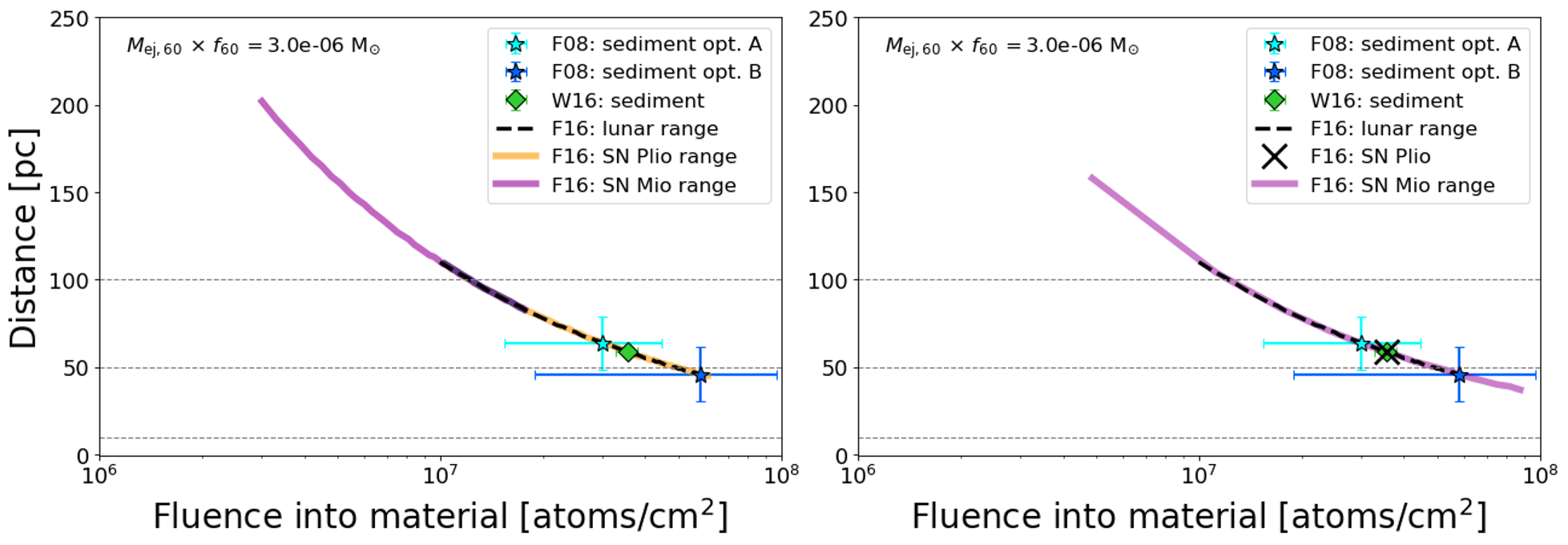}
  \caption{\textit{Lunar fluence portioning}. \textbf{Left:} Lunar fluence portioned with the W21 fluence ratios (see Subsection~\ref{ssec:lpw21}), along with the F16 published full lunar range.  SN Plio (3.1 Mya) is in gold and SN Mio (7 Mya) is in purple; note the overlap.  SN Plio is 90\% of the lunar \fe60 and therefore traces nearly the same range as the originally published data (dashed line).  \textbf{Right:} Lunar fluence portioned assuming the W16 sediment data is the full fluence for SN Plio (see Subsection~\ref{ssec:lpw16}).  The SN Plio fluence is plotted with a black X, directly on top of the W16 sediment point.  The purple line shows the possible remaining fluence and distance range for SN Mio, along with the original lunar data as a dashed line.  Included on both plots for demonstration are the F08 and W16 sediment data.}
  \label{fig:dist_lunar}
\end{figure}

Using $\mej=3.0\times10^{-5}$ \msol, $\df = 0.1$, and $\trav = 0.1$ Myr, we find that SN Plio occurs between $45 - 110$ pc from Earth and SN Mio occurs between $80 - 200$ pc.  The left plot in Fig.~\ref{fig:dist_lunar} shows these two ranges in gold and purple, respectively, along with the original lunar range quoted in F16.  Note that the lunar fluence for SN Plio is about 10\% less than the total lunar fluence for the last 8 Myr quoted in F16; the 10\% loss results in a distance that is only a few pc different, meaning that a 10\% difference in fluence does not have a large impact on the distance calculation.
However, this difference does allow us to extract additional information about the distance to the second supernova.

The full range of the lunar fluence corresponds to a distance range for SN Plio from $45 - 110$ pc.  It is interesting to note that the W16 sediment data (assumed to have a 100\% uptake factor) falls exactly on this band and the two possibilities for the F08 sediment data include the band within error.  These are completely independent measurements and in fact occur on separate bodies in the solar system.  An extension of this work is to use the W16 sediment fluence and calculated distance to pinpoint what the actual lunar fluence is for SN Plio, with the remainder of the excess \fe60 then originating from SN Mio, which we do below.  

\newpage
\subsubsection{Data-driven portioning with W16 results} \label{ssec:lpw16}

As noted in the previous section, the W16 sediment fluence falls exactly in the range of the lunar fluence.  In this section, we make the assumption that the W16 sediment \textit{is} the fluence from SN Plio at 3 Mya; therefore, any remaining lunar \fe60 fluence detected is from SN Mio at 7 Mya.  Once again undoing the decay correction on the lunar fluence and the W16 sediment fluence with Eq.~(\ref{eq:decaycor}), we can subtract the ``dug up'' W16 fluence from the lunar fluence, redo the 7 Mya decay correction, and recalculate the possible distance range to SN Mio with Eq.~(\ref{eq:dist}).  Using the same astrophysical parameters ($\mej=3.0\times10^{-5}\msol$, $\df = 0.1$, and $\trav = 0.1$ Myr), we find that the SN Mio distance range with this method is $40 - 160$ pc. 

The right plot in Fig.~\ref{fig:dist_lunar} shows the results of this calculation.  The fluence from SN Plio is denoted with a black X and is plotted directly over the W16 sediment fluence (as these are the same number).  The shaded purple line represents the full possible range of fluence and distance for SN Mio, with the original F16 range plotted as a black dashed line. 

\section{Models}\label{sec:models}

The \fe60 signal found in the natural archives on Earth can be used to find the \fobs, \tarr, and \up\ parameters needed to calculate the supernova distance in Eq.~(\ref{eq:dist}).  For the remaining three parameters of \mej, \df, and \trav, we turn to astrophysical models and observations to provide additional constraints, and we explore the allowed ranges that remain.  A brief summary of the parameter ranges are listed in Tab.~\ref{tab:astro_param} and these ranges are discussed in detail in the following subsections.

\begin{table}[ht]
    \centering
    \caption{Astronomical Parameter Ranges}
    \begin{tabular}{lc|c}
    \hline \hline
\multicolumn{2}{c|}{Parameter} & Range \\
\hline
Ejected \fe60 mass, \mej & [\msol] & $3\times10^{-6} - 3\times10^{-4}$\\
Dust fraction, \df & & $1\% - 100\% $\\
Travel time, \trav & [Myr] & $0.1 - 1.5$\\
\hline \hline
    \end{tabular}
        \vspace{3mm}
    \label{tab:astro_param} \\
\end{table}

\subsection{Ejected Mass} \label{ssec:mej}

There are no available measurements of \fe60 yields in individual supernovae.
Thus we have two options.  One is to rely on theoretical predictions. 
The other is to use observations of \fe60 gamma-ray emission from the Galaxy to find an average \fe60 yield.  We consider each of these in turn.

\subsubsection{Supernova Calculations of \texorpdfstring{\fe60}{60Fe} Yields}

Finding the ejected mass of \fe60 from a supernova requires modeling explosive nucleosynthesis in the shell layers of the progenitor as it explodes.  The \fe60 is not made in the core of the CCSN but instead from neutron capture onto pre-existing iron in the shell layers (for a recent review, see \citealp{diehl2021}).  For this reason, we exclude explosion models with low metallicity or which do not track nucleosynthesis in the shell layers, such as the \citet{curtis2019} ``s model'' and the \citet{wanajo2018} ``s models''.  

\begin{figure}[htb]
	\centering
  \includegraphics[width=18cm]{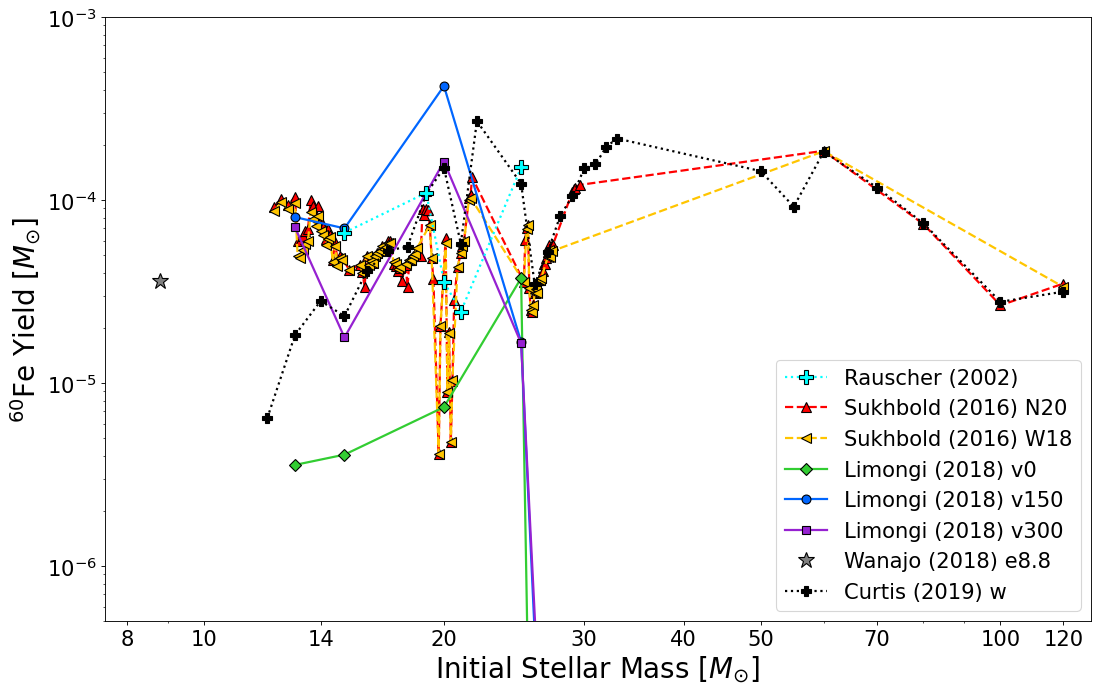} 
  \caption{\textit{Supernova model yields for \fe60}. Masses within models are connected with lines.  Note that the possible ranges for \mej\ vary between about $3\times10^{-6}$ and $3\times10^{-4}$ \msol.  The three \citet{limongi2018} models drop off the plot and demonstrate direct collapse to a black hole.  The relevant paper citations and details can be found in Tab.~\ref{tab:yields}.}
  \label{fig:sn_yields}
\end{figure}

\begin{table}[ht]
    \centering
    \caption{SN \fe60 Yield Models}
    \begin{tabular}{lc|cccc}
    \hline \hline
Authors & Model & Metallicity & Rotation & Mass range (\msol) & Type\\
\hline \hline
\citet{rauscher2002} &  & solar & no & 12-25 & CCSN\\
\hline
\citet{sukhbold2016} & N60 & 1/3 solar & no & 12.25-120 & CCSN\\
\citet{sukhbold2016} & W18 & 1/3 solar & yes & 12.25-120 & CCSN\\
\hline
\citet{limongi2018} & v0 & solar & no & 13-120 & CCSN\\
\citet{limongi2018} & v150 & solar & yes & 13-120 & CCSN\\
\citet{limongi2018} & v300 & solar & yes & 13-120 & CCSN\\
\hline
\citet{wanajo2018}\footnote{The e8.8 model used in \citet{wanajo2018} is the same as used in \citet{wanajo2013}.} & e8.8 & solar & no & 8.8 & ECSN\\
\hline
\citet{curtis2019} & w & solar & no & 12-120 & CCSN\\
\hline \hline         
    \end{tabular}
        \vspace{-1.0mm}
    \label{tab:yields} 
\end{table}

There are some additional constraints we can place on the available nucleosynthesis models.  As discussed in \citet{fry2015}, the supernova must be a CCSN or ECSN in order to produce sufficient \fe60. It must also be close enough to Earth for its debris to reach the solar system.  While we have already excluded models with low metallicity which will not make \fe60, the progenitor should already be at or near solar metallicity due to its proximity to Earth and the time of the explosion ($\lesssim$ 10 Mya). 

Table~\ref{tab:yields} highlights the relevant simulation parameters for the selected models.  We focus on four recent publications which model stars of solar metallicity and include 
\fe60 production in their nucleosynthesis reactions \citep{sukhbold2016,limongi2018,wanajo2018,curtis2019}.  \citet{rauscher2002} is included to enable comparison with the data used in \citet{fry2015}.  Figure~\ref{fig:sn_yields} then plots the \fe60 yields from the five different groups, with each individual model plotted separately.  Lines connect the individual masses to give a better sense of the model's range; the single point from \citet{wanajo2018} focuses on a specific mass ECSN model.  

\begin{figure}
    \centering
    \includegraphics[width=0.8\textwidth]{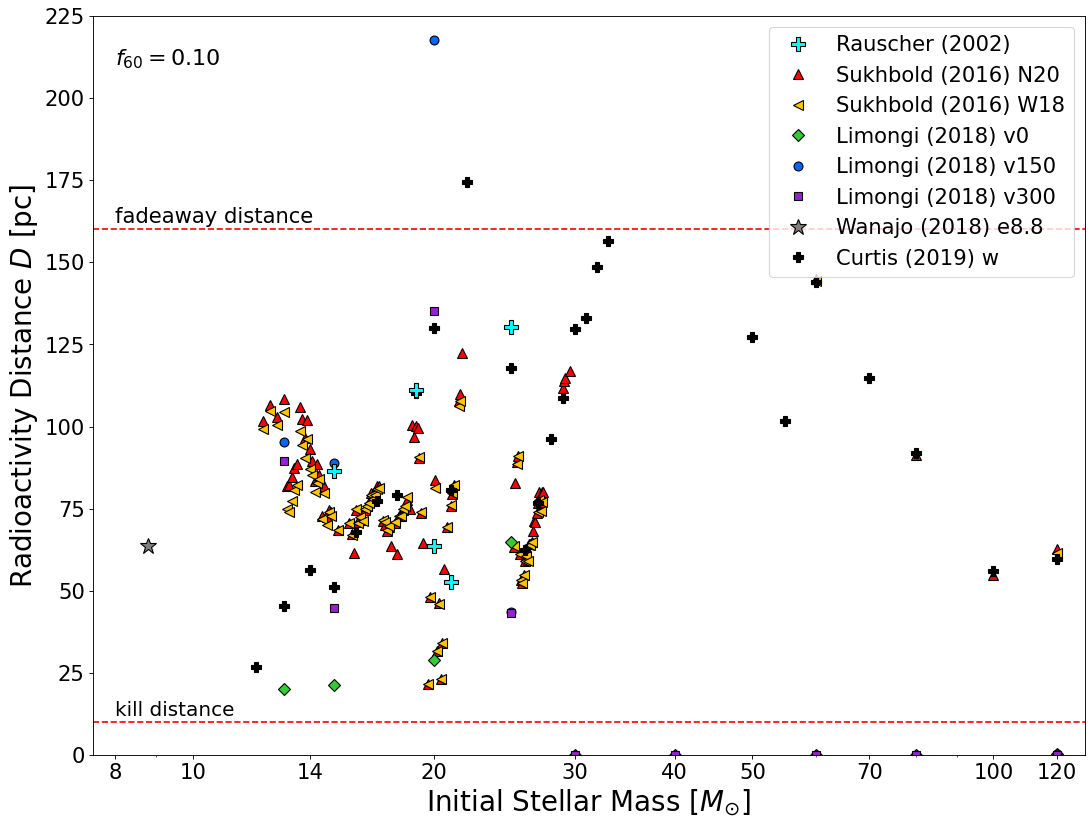}
    \caption{Radioactivity distance for the core-collapse \fe60 yields shown in Fig.~\ref{fig:sn_yields}.  
    Results use the W16 sediment data and assume $f_{60}=0.1$.  Estimates of limiting distances are shown as dashed horizontal red lines: the kill distance at 10 pc is a lower limit, and the fadeaway distance at 160 pc is an upper limit \citep{draine2011}.  We see that for most models, the radioactivity distance lies between these limits, showing that a consistent picture is possible for this wide class of supernova models.}
    \label{fig:D_rad_vs_yields}
\end{figure}

The focus of this paper is not to describe these supernova models in detail, but instead to find a range of the ejected mass of \fe60 that can be used in Eq.~(\ref{eq:dist}).  From Fig.~\ref{fig:sn_yields}, we see that the possible range extends from $3\times10^{-6} - 3\times10^{-4}$ \msol\ and is covered fairly evenly by all groups.  A recent further discussion of \fe60 production in supernovae appears in \citet{diehl2021}.

Figure \ref{fig:D_rad_vs_yields} shows the radioactivity distance implied by these yields.  Here we adopt the \fe60 fluence from W16 sediments, along with a dust fraction \df$=0.1$.  We see that the wide range of yields in Fig.~\ref{fig:sn_yields} leads to a substantial range in the radioactivity distance, even with the \mej$^{1/2}$ scaling.  Encouragingly, we see that almost all models give distance between these limits.  This is represents a nontrivial success of the nearby supernova scenario, because calculation of $D$ in eq.~(\ref{eq:dist}) depends only on the fluence measurements and yields, with no astrophysical distances built in.  Moreover, the allowed distance range encompasses the Local Bubble, and star clusters proposed to be the sites of the supernovae, as discussed below in Section~\ref{sec:discussion}.  

While Fig.~\ref{fig:D_rad_vs_yields} shows the full range of masses for which \fe60 is presented in recent models, these stars are not all equally probability.  The stellar initial mass function shape indicated that lower mass CCSN progenitors should common; here we see that across several models these all provide reasonable distances.  Indeed we do not see clear systematic differences between the distances inferred with lower vs higher mass models, reflecting the lack of a clear trend in \fe60 yields vs progenitor mass in Fig.~\ref{fig:sn_yields}.  This suggests that it will be difficult or impossible to use \fe60 alone to probe the mass of the progenior; for this multiple SN radioisotopes are needed.

We stress that the distances shown in Fig.~\ref{fig:D_rad_vs_yields} derive from the dust fraction choice $f_{60} = 0.1$, and scale as $D \propto f_{60}^{1/2}$.  Thus, significant systematic changes in $D$ can result from different choices for this poorly-determined parameter.  This point is discussed further in the following section.

With this caveat in mind, it is notable that Figure \ref{fig:D_rad_vs_yields} also shows that a few \fe60 calculations do not fall into the allowed range. Most notable are the \citet{limongi2018} models, which give $D \approx 0$ for progenitor masses $\ge 30 M_\odot$.  This arises because in these models there is a direct collapse to a black hole without an explosion and the accompanying ejection of nucleosynthesis product; thus the only \fe60 that escapes is the small amount in the stellar wind.  Clearly these models are excluded, but the lower mass \citet{limongi2018} give \fe60 in good agreement with other calculations and thus give plausible radioactivity distance.

Finally, we note that Fig.~\ref{fig:D_rad_vs_yields} updates a similar calculation by \citet[][their Fig.~3]{fry2015}.  Our results are broadly quite similar.  This agreement is somewhat accidental, since those earlier results were based on K04 and F08 \fe60 crust measurements prior to the more reliable sediment measurement of W16 (used in this work).  Moreover, uptake and dust assumptions were different.  On the other hand, while detailed stellar yields have changed, they continue to span a similar range.

\subsubsection{The Average \texorpdfstring{\fe60}{60Fe} Yield from Gamma-Ray Line Observations}

Gamma-ray line astronomy provides an estimate of the average \fe60 yield from supernovae.  The radioactive decay of \fe60 atoms leads to emission of gamma-ray and X-ray lines.  Interstellar \fe60 thus produces observable gamma-ray lines that probe its production over approximately one \fe60 lifetime.

Measurements of diffuse Galactic gamma rays give a steady state \fe60 mass of $M_{60,\rm SS} = 2.9^{+2.2}_{-0.78} $ \msol\ \citep{diehl2021}.  In steady state, $M_{60,\rm SS} = \avg{\mej} \, \tau_{60}\,{\cal R}_{\rm CCSN}$, where ${\cal R}_{\rm CCSN} = 1.79 \pm 0.55 \ {\rm century}^{-1}$ is the present-day Galactic core-collapse rate \citep{rozwadowska2021}.  We thus find the mean \fe60 yield to be
\begin{eqnarray}
\label{eq:Mejavg}
    \avg{\mej} & = & \frac{M_{60,\rm SS}}{\tau_{60} \, {\cal R}_{\rm SN}}  \nonumber  \\
    & = & 4^{+4}_{-2} \times 10^{-5} M_\odot \ \pfrac{M_{60,\rm SS}}{2.85 \ \msol} \pfrac{1.79 \ \rm event/century}{{\cal R}_{\rm SN}} \ \ .
\end{eqnarray}
\noindent This result averages over all core-collapse supernovae, which are assumed to be the only important \fe60 source.  If another source such as AGB stars makes an important contribution to the Galactic \fe60 inventory, then the mean SN yield would be lower.

Interestingly, the result in Eq.~(\ref{eq:Mejavg}) is in the heart of the predictions shown in Fig.~\ref{fig:sn_yields}.  This supports the idea that core-collapse supernovae indeed dominate \fe60 production, and suggests that the theoretical predictions are in the right ballpark.  Further, if this is a typical yield, then there are implications for the dust fraction, 
to which we now turn.

\subsection{Dust Fraction} \label{ssec:dustfrac}

Our current understanding of the interactions between the SN shock front and the heliosphere require the \fe60 to be in the form of dust grains in order to reach Earth \citep{benitez2002,athanassiadou2011,fry2015,fry2016}.  The heliosphere blocks the supernova shock front from pushing inward past 5 au for supernova distances $> 30$ pc \citep{miller2022}, 
and therefore only dust grains $\gtrsim 0.1 \mu$m can ballistically push through the barrier \citep{athanassiadou2011,fry2020}.  The dust fraction parameter \df\ describes the fraction of ejected \fe60 mass which condenses into dust and therefore makes it to Earth --- this parameter encompasses the multiple boundaries and survivability filters that the dust must traverse.  As laid out in \citet{fry2015} Tab.~3, these include:

\begin{itemize}
    \item[1.] The amount of \fe60 that initially forms into dust.
    \item[2.] The amount of \fe60-bearing dust which survives the reverse shock, sputtering, collisions, and drag forces in the supernova remnant (SNR) to then encounter the heliosphere.
    \item[3.] The amount of dust that makes it past the shock-shock collision at the heliosphere boundary.
    \item[4.] The amount of dust which manages to traverse the solar system to 1 au and be collected on Earth (and the Moon).
\end{itemize}

Observational work on SN1987A has demonstrated refractory elements form dust within years of the explosion and that nearly 100\% of supernova-produced elemental iron condenses immediately into dust \citep{matsuura2011, matsuura2017,matsuura2019,cigan2019,dwek2015}.  The composition of \fe60-bearing dust is of specific interest, especially considering that different compositions have significantly different survival rates due to grain sputtering \citep{fry2015,silvia2010,silvia2012}.  Given that the \fe60 is formed in the shell layers of the progenitor and not in the iron core with the bulk of the supernova-produced elemental iron \citep{diehl2021}, it is quite possible that the \fe60 dust is not in predominately metallic iron grains --- this in turn can affect the dust's survival chances \citep{silvia2010}.

The degree to which dust is produced and destroyed within supernova remnants is an area of ongoing research; a recent review by \citet{micelotta2018} summarizes the current theory and observational work.  CCSN are producers of dust, as shown by observations of grain emission in young remnants, e.g., in recent JWST observations \citep{shahbandeh2023}. The portion of grains that survive and escape the remnant is more difficult to establish.  Factors such as dust composition, size, clumpiness within the remnant, and the density of the ambient medium all impact the dust survival rate. Table~2 in \citet{micelotta2018} lists the calculated dust survival fractions within the SNR for various models and simulations.  These dust fractions vary wildly between models and  ultimately range from 0\% to 100\% of the supernova-produced dust surviving the forward and reverse shocks.  More recent papers continue to find that a large range is possible \citep{slavin2020a,marassi2019}. 

There are many dynamics and effects within the solar system which can filter dust grain sizes and prevent grains from easily entering the inner solar system \citep{altobelli2005,mann2010,wallner2016,altobelli2016,strub2019}; however, the $\sim 100$ km/s speed at which the supernova-produced dust is traveling causes these effects to be negligible \citep{athanassiadou2011, fry2015}.  Therefore, we will consider the dust fraction at Earth's orbit to be equal to the surviving dust fraction within the SNR.  Note that this assumption is in contrast to the assumptions made in Tab.~2 in \citet{fry2015}, which assumes that only 10\% of the metallic iron (Fe) dust and none of the troilite (FeS) dust cross the heliosphere boundary.\footnote{Upon closer examination of the sources cited \citep{linde2000,silvia2010,slavin2010}, we find that they are focused on ISM grains which are traveling $\sim 26$ km/s; our SNR grains are traveling at $\sim$100 km/s.  Further research is needed to work out the details of the dust's ability to penetrate the heliosphere depending on size and velocity --- for the purpose of this paper, we follow the approach outlined in W16.}  In contrast, \citet{wallner2016} follows a similar route to our approach in this paper, in assuming that the \fe60 dust survives from the shock boundary to the Earth essentially unchanged.

With the assumptions that 100\% of the ejected mass of \fe60 condenses into dust and that the dust survives the heliosphere boundary fully intact, the main source of dust loss is within the SNR itself.  As discussed above, the fraction of dust that survives the SNR environment ranges wildly depending on model parameters such as size, composition, and ambient environment --- although at least some of the dust must survive this journey, as we find the \fe60 signal on Earth.  For the purposes of this work, we do not focus on the details of dust survival calculations but instead choose a general dust fraction of between $1-100\%$.

It should be noted that W16 and W21 calculate the dust survival fraction using the fluence from the W16 sediment sample (which samples 100\% of the \fe60 flux onto Earth).  They assume that the \fe60 dust traverses the solar system essentially unchanged and therefore the \fe60 fluence at Earth is the same as the interstellar fluence.  W16 finds a dust survival fraction of $0.4-9\%$, using the additional assumptions that source of the 3 Mya signal occurred somewhere between 50 and 120 pc from Earth and with an ejected mass of $\mej \sim 9\times10^{-5}$ \msol.\footnote{This \mej\ value is found by assuming the supernova signal at 3 Mya is the debris from three separate supernovae, given the long $> 1$ Myr infall timescale (similar to what is proposed in \citealp{breitschwerdt2016}). As discussed in \citet{ertel2023}, \citet{fry2020}, and \citet{chaikin2022}, more than one SN is not required to produce the observed signal and therefore this might overestimate the ejected mass.  However, the value is still well within the range of possible values for \mej\ for a single supernova, as seen in Subsection~\ref{ssec:mej}.}  We do not use these numbers directly in this work, as they are unavoidably correlated to the observed fluence.  
Thus we see that our adopted benchmark value $\df = 10\%$ lies comfortably within the large allowed range, but clearly more work is needed to understand this quantity better.  Indeed, as W16 and W21 show and we will discuss below, \fe60 observations place novel limits on dust survival.

\subsection{Travel Time} \label{ssec:traveltime}

The travel time parameter (\trav) is defined as how long the supernova dust containing the \fe60 will travel within the remnant before reaching Earth --- specifically, it is the dust's travel time before the start of the \fe60 signal.\footnote{The dust containing the \fe60 will continue to raindown for $\gtrsim$ 1 Myr after the initial deposition.  We therefore define the travel time as tracing the SNR shock front and not the specific dust dynamics within the remnant that extend the signal.}  
As the remnant expands, it cools and slows, achieving a maximum size of around $100-200$ pc \citep{fry2015,micelotta2018}.  At most, this expansion should last around $1-2$ Myr (see Fig.~1 in \citealp{micelotta2018}).  We therefore consider the range for \trav\ to be $0.1 - 1.5$ Myr.  Note that any value of \trav\ less than the half-life of \fe60 ($t_{1/2}$ = 2.62 Myr) has little to no impact on the resulting distance calculated.  

It should be noted that at least SN Plio exploded in the Local Bubble environment and not the general ISM (\citealp{zucker2022}, for further discussion, see Subsection~\ref{ssec:lb_imp}).  The low density does not change our estimate of \trav:  while the low density allows the remnant to expand farther, the blast also travels faster and therefore this extra distance is negligible. 

There are also additional constraints due to the fact that we have observed an \fe60 signal on Earth twice in the last 10 Myr.  The dust containing \fe60 arrived on Earth approximately 3 Mya for SN Plio and 7 Mya for SN Mio.  Our ability to detect the \fe60 signal from that point is dependent on the half-life of \fe60.  For SN Plio, we have lost about one half-life of the isotope since deposition, and about three half-lives for SN Mio.  It is therefore reasonable to believe that the dust is not traveling for millions of years before reaching Earth, as further loss of \fe60 would imply either enormous \fe60 supernova production or raise questions over our ability to detect the \fe60 signal at all.

\section{Summary of Results}\label{sec:results}

\begin{figure}
	\centering
  \includegraphics[width=12cm]{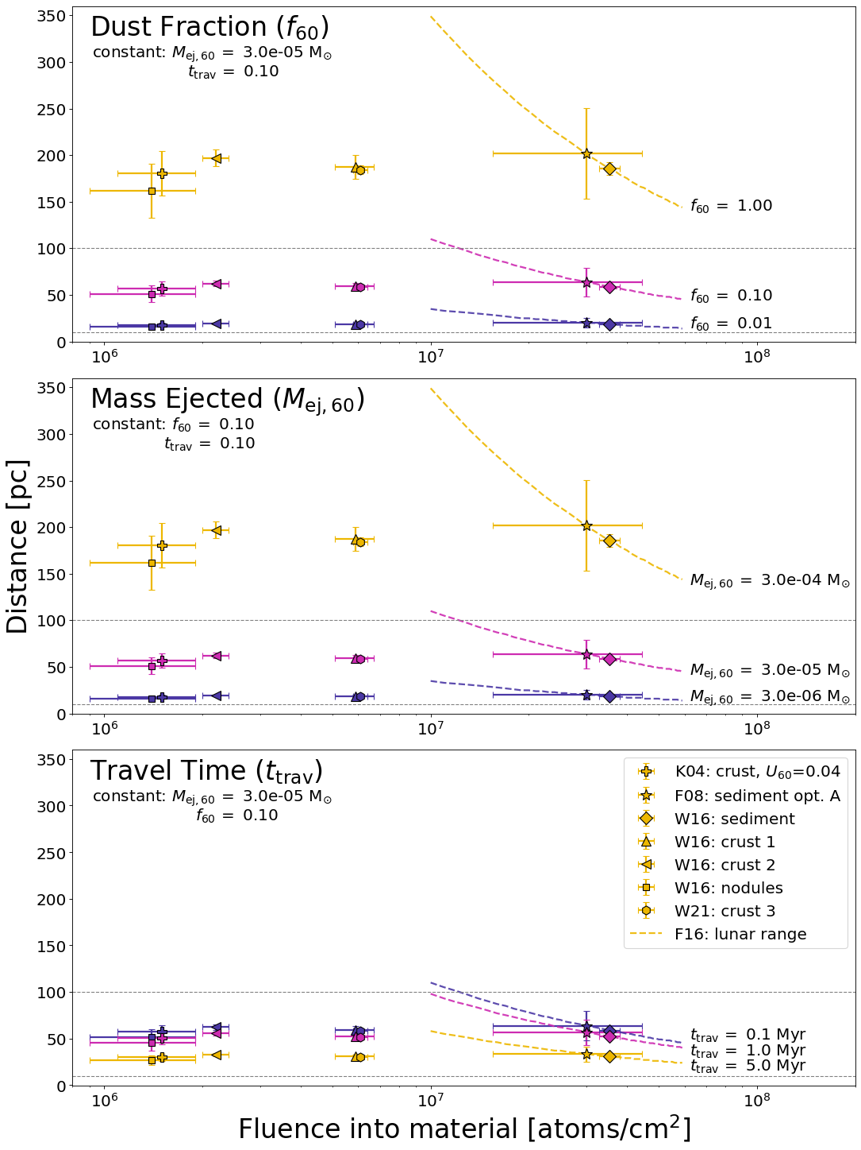}
  \caption{\textit{Effect of the three astrophysical parameters on the distance to SN Plio}. For the purposes of reducing visual confusion, we only show the K04 distance calculated with $\up = 0.04$ and the F08 option A points.  Note that this means the F08 and F16 distances are the only ones not correlated to the W16 sediment. \textbf{Top:} Dust fractions (\df) of 1\%, 10\%, and 100\%, holding the ejected mass constant and travel time constant. \textbf{Middle:} Ejected masses (\mej) of $3.0 \times 10^{-6}$ to  $3.0 \times 10^{-4}$ \msol\, holding the dust fraction and travel time constant.  \textbf{Bottom:} Travel times (\trav) of 0.1, 1.0, and 5.0 Myr, holding the dust fraction and ejected mass constant.  The dark blue, magenta, and yellow points for each plot indicate the distances calculated with the low, middle, and high values for each range, respectively.  Lines have been drawn at 10 pc and 100 pc.}
  \label{fig:comp}
\end{figure}

We combine the ranges of the three astronomical parameters discussed in Section~\ref{sec:models} with the observed \fe60 signal in Section~\ref{sec:data} to calculate the full possible range of distances to SN Plio (3 Mya).  Figure~\ref{fig:comp} maps out these results, following the same plotting convention of distance to supernova vs fluence into the material used in Figs.~\ref{fig:dist} and~\ref{fig:dist_lunar}.  For each subplot, two of the parameters are held constant at a mid-range value while the third is allowed to vary for the full range considered in this paper.  

To reduce visual confusion, we have chosen to only show the F08 option A and the K04 crust with $\up=0.04$ points; using the K04 crust with $\up=0.04$ means that the F08 and F16 distances are the only ones not correlated to the W16 sediment (see Subsection~\ref{ssec:knie}).  As a reminder, the error bars on the fluence reflect the actual fluence errors --- the error bars on the distance only trace the fluence errors and do not account for the error in \up\ present on the FeMn crusts and nodules (see Section~\ref{sec:data}).

\begin{table}[htbp]
    \centering
    \caption{Influence of \astropar\ on SN Plio distance}
    \begin{tabular}{c|c|c}
    \hline \hline
    \astropar\ [\msol] & Approximate distance [pc] & Notes \\
\hline
$3.0\times10^{-8}$ & $5.87 \pm 0.22 $ & lowest combination\\
$8.85\times10^{-8}$ & $10 $ & 10 pc supernova\\
$3.5\times10^{-7}$ & $20 $ & 20 pc supernova\\
$2.18\times10^{-6} $ & $50 $ & 50 pc supernova \\
$3.0\times10^{-6} $ & $59.7 \pm 4.0 $ & mid-range \\
$8.75\times10^{-6}$ & $100 $ & 100 pc supernova\\
$3.0\times10^{-4}$ & $586 \pm 22$ & highest combination \\
\hline \hline
    \end{tabular}
        \vspace{3mm}
    \label{tab:dist_influence} \\
    \parbox{0.9\textwidth}{\textbf{Caption:} \astropar\ ranges from the lowest combination ($\mej=3.0\times10^{-6}$ \msol, $\df=1\%$) to the highest combination ($\mej=3.0\times10^{-4}$ \msol, $\df=100\%$).  $\astropar = 3.0\times10^{-6}$ \msol\ reflect a mix of the highest and lowest parameter combination and also covers the mid-range for both combinations. Specific distances of interest (10, 20, 50, 100 pc) are also singled out.  Distance is calculated using the W16 sediment fluence; the error on the distance reflects the observed fluence error.}
\end{table}

As can been seen in the bottom plot of Fig.~\ref{fig:comp}, the travel time does not have a large effect on the distance range and is the least important parameter.  The difference between 0.1 and 1.0 Myr is negligible, on the order of $\sim$ 5 pc, which is well within the uncertainties of the Earth-based parameters.  It is only when the travel time is increased to a non-physical 5 Myr that any effect is observed, and that effect is minimal compared to the ranges seen in the \mej\ and \df\ parameters. 

Both the dust fraction and ejected mass range two full orders of magnitude and have a significant influence on the distance.  Untangling these parameters is challenging, and as discussed in Section~\ref{sec:math}, the product \astropar\ is more robust.  Table~\ref{tab:dist_influence} lists a range of possible \astropar\ values, covering the lowest and highest combinations, the middle of both ranges, as well as four distances of specific interest.  All of the approximate distances in Tab.~\ref{tab:dist_influence} are calculated using the W16 sediment fluence, as we believe this measurement best reflects the \fe60 fluence from SN Plio (see Subsection~\ref{ssec:wallner} for further details).

Both the lowest and highest possible combinations of \mej\ and \df\ put SN Plio at implausible distances:  any distance closer than 20 pc should have left distinct biological damage tracers in the fossil record \citep{melott2011,fields2020}, while distances farther than $\sim$ 160 pc prevent the dust from reaching Earth \citep{fry2015}.  The remaining combinations produce a large range of possible supernova distances.  Using an $\astropar$ range of $5\times10^{-7} - 5\times10^{-5}$ \msol\ yields distances between $20 - 100$ pc; these values for \astropar\ can be found using any high-low or low-high combination of \mej\ and \df\ as well as the middle range for both parameters.

In the interest of a complete analysis, we have expanded \mej\ and \df\ to the full possible range of values --- this does not necessarily reflect the most likely range.  While the available \fe60 yield models cover the full $3\times10^{-6} - 3\times10^{-4}$ \msol\ for the mass range of interest ($8 - 30$ \msol), as seen in Fig.~\ref{fig:sn_yields}, the dust fraction survival range is more likely to be between $1-50$\% \citep{wallner2016,wallner2021,slavin2020a}.  

\subsection{Distance to the 7 Mya supernova}

The W21 dataset includes a distinct FeMn crust measurement for the 7 Mya SN.  We can use this fluence to calculate the distance to SN Mio following the same procedure as described for SN Plio.  The fluence for SN Mio is $\mathcal{F} = 1.77\pm0.25\times 10^6$ atoms/cm$^2$, with an uptake factor of $\up=17\%$ \citep{wallner2021} --- the fluence is already decay corrected, so the arrival time is not needed.  We assume the same average supernova properties as SN Plio, with $\mej = 3.0\times10^{-5}$ \msol\ and $\df=0.1$, although these properties do not have to be the same for both supernovae.  Using Eq.~(\ref{eq:dist}), we find that the distance to SN Mio is
\begin{equation} \label{eq:mio_dist}
    D(\mbox{SN Mio}) \simeq \left( 108 \pm 8 \ \rm{pc.} \right) 
    \pfrac{\df}{0.1}^{1/2}
    \pfrac{\mej}{3.0\times10^{-5} \msol}^{1/2} 
\end{equation} 
where the {\em errors reflect only the reported uncertainties in the fluence}.
Calculating the distance using a ratio of the fluences found in W21 provides the same results, with $D_2 = D_1 \sqrt{\mathcal{F}_1/\mathcal{F}_2} \sim 110$ pc.  Similarly, we note that the W21 crust 3 fluence (for SN Mio) and the distance calculated in Eq.~(\ref{eq:mio_dist}) fall exactly on the estimated range for SN Mio as shown in the lunar fluence (see Subsection~\ref{ssec:lpw21}) --- this is unsurprising, as all three of these calculations are based off of the W21 crust fluences and are therefore correlated.

\begin{figure} 
	\centering
  \includegraphics[width=12cm]{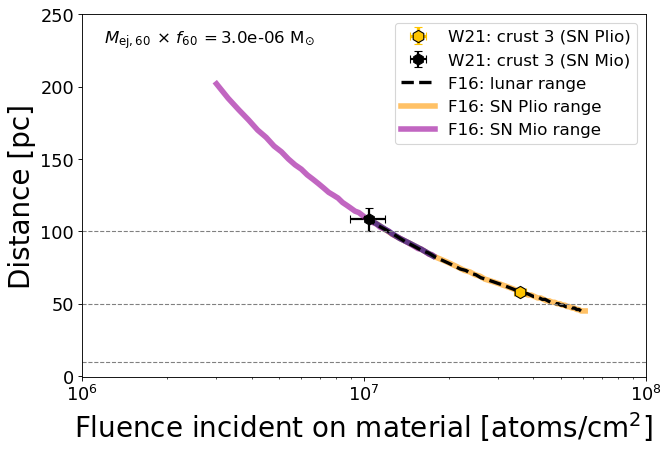}
  \caption{\textit{Distance to SN Mio.}  Plotted are the portioned lunar fluences with gold (SN Plio) and purple (SN Mio) lines (see Subsection~\ref{ssec:lpw21} and Fig.~\ref{fig:dist_lunar}), the original lunar fluence with a black dashed line, the W21 crust 3 fluence for SN Plio in yellow, and the W21 crust 3 fluence for SN Mio in black.  For ease of comparison, the W21 crust 3 measurements have been corrected for the uptake factor of 17\%; they are therefore directly equivalent to the lunar measurements, which assume a 100\% uptake.  The distances to both SN Plio and SN Mio are calculated using the same \astropar.}
  \label{fig:sn_mio}
\end{figure}

Figure~\ref{fig:sn_mio} shows the distance to SN Mio compared to SN Plio.  The distance calculated with the W21 crust 3 fluence for SN Mio is plotted along with the portioned lunar range corresponding to SN Mio.  Two equivalent values are shown for SN Plio:  the W21 crust 3 fluence for SN Plio and the portioned lunar range for SN Plio.  To make the W21 crust 3 data directly comparable to the F16 lunar fluence, we have corrected for the 17\% uptake factor in the crusts.\footnote{As a reminder, the uptake factor accounts for how much of the \fe60 fluence is lost between arrival at Earth and absorption into the sampled material.  The lunar regolith, similar to the deep-sea sediments, has a 100\% uptake factor, as it is assumed no fluence is lost.  The W21 crust has a 17\% uptake factor (see Section~\ref{sec:data}).  Thus in Fig.~\ref{fig:sn_mio}, we have accounted for the remaining 83\% of the fluence in the crust measurements, to make them directly comparable to the lunar regolith measurements.} 

We note that, given the available measurements, the fluence for SN Mio is around a factor of two smaller than the fluence for SN Plio.  These values have been decay-corrected and therefore this difference is real, excluding unknown geophysical parameters that could affect the signals.  As discussed in Section~\ref{sec:models}, the influence this has on the calculated distance could be the result of a genuine difference in distances between the two supernovae or due to differences in the \astropar\ parameter; in reality, it is likely a combination of both scenarios.    

Ideally, this calculation should be repeated with greater precision when there are more data available on SN Mio.  With a detailed time profile, such as can be provided with sediment measurements, we might be able to investigate differences in the astronomical properties between the two supernovae; although as the above sections discuss there are significant ranges in such factors as the \fe60 ejected mass and dust fraction, and differences in the supernovae distances could obscure these variations.  

\section{Discussion}\label{sec:discussion}

Our results have an interplay with several areas of astrophysics, which we summarize here.

\subsection{External limits on distance:}

As discussed in Section~\ref{sec:results}, it is difficult to limit the possible distance range to SN Plio from the astrophysical parameters alone.  The theoretical values for both the ejected mass and dust fraction of \fe60 individually cover two orders of magnitude and, although we can rule out a few of the most extreme combinations, we are unable to use them to put strong limits on the supernova distance.  We therefore turn to other methods that can be used to provide limits.  

\textbf{Biological damage and extinctions:}

There has been a long history of speculating on the damage that supernova-driven cosmic rays could do to ozone in the atmosphere and implications for life on Earth \citep{shindewolf1954,shklovskii1966,ruderman1974,benitez2002,melott2011,thomas2016}.  \citet{gehrels2003} was the first to calculate a ``kill distance'' of 8 pc, inside of which supernovae could be responsible for mass-extinction events --- by depleting the ozone, UVB radiation from the Sun can cause significant damage to DNA.  Later updates by \citet{fry2015} extend the kill distance to 10 pc which we have adopted in this paper, although recent work suggests that the distance could be as far as 20 pc \citep{thomas2023}.  As there is still life on Earth, we can rule out distances $< 10$ pc.

While more minor biological damage events and other climate-driven disruptions are still possible \citep{thomas2018,melott2019,melott2019a}, we can also rule out any distance that would result in a mass-extinction, as there are no major mass-extinction events 3 Mya.\footnote{There are also no such events 7 Mya, for SN Mio.}  This prevents the distance to SN Plio from being within 20 pc of Earth \citep{fields2020}.   

\textbf{Cosmic ray distance calculation:}

There is a local \fe60 component measured in cosmic rays, which is associated with a nearby supernova around 2 Mya \citep{binns2016}.  \citet{kachelriess2015,kachelriess2018} examines the proton, antiproton, and positron fluxes and anomalies in the cosmic ray spectrum; they argue for a local source in addition to the contribution from cosmic ray acceleration in supernova remnants throughout the Galaxy.  \citet{savchenko2015} shows that a feature in cosmic ray anisotropy at $2-20$ TeV is due to a single, recent, local event.  Using these data they calculate a distance, estimating the source to be roughly 200 pc from Earth.  It should be noted that while the \fe60 cosmic ray background is local and recent, these analysis relied on the detection of the \fe60 signal for SN Plio --- we now know of two supernovae which occurred in the last 10 Myr and further analysis of the local cosmic ray background will need to take this into account. 

\textbf{Nearby clusters:}

The massive stars that explode in CCSN or ECSN tend to form in clusters, most likely including the two near-Earth supernovae.  We can use the locations of nearby stellar groups and associations as another method for constraining the supernova distances.  The Tuc-Hor association is $\sim 60$ pc from Earth \citep{mamajek2015} and considered to be the mostly likely to host SN Plio \citep{fry2015,hyde2018}.\footnote{As a reminder, SN Mio was only measured in detail in 2021 \citep{wallner2021} and is therefore not considered in these evaluations, although we expect it to have similar results.}  \citet{hyde2018} surveyed the local associations and groups within 100 pc and concluded that Tuc-Hor is the only association with an initial mass function (IMF) large enough to host a CCSN, thus making it the most likely candidate.

Another consideration is the OB Sco-Cen association, favorable because it has the IMF to host multiple CCSN and is likely the source of the supernovae which formed the Local Bubble \citep{frisch2011,fry2015,breitschwerdt2016}.  Sco-Cen is located around 130 pc from Earth \citep{fuchs2006} and this puts it at the farther edge of the possible distance range.  \citet{neuhauser2020} even backtracked the positions of both a nearby pulsar and runaway star, claiming that they shared a common binary in proximity to Sco-Cen 1.78 Mya and could be the remnants of SN Plio --- unfortunately, the timing does not quite work out for SN Plio, as the initial \fe60 signal starts at 3 Mya and the progenitor of the signal must precede the deposition time.

\textbf{Dust stopping time:}

\citet{fry2020} simulates supernova-produced dust under the assumption that the dust is charged and will therefore be confined within the magnetized remnant.  They find that their models can consistently propagate grains to 50 pc, but that greater distances are affected by magnetic fields and drag forces and unlikely to be reached.  Although further research is needed, this work provides an interesting potential limit on the distances to SN Plio and SN Mio that is not dependent on the local interstellar low-density environment.

\subsection{Local Bubble implications} \label{ssec:lb_imp}

An interesting complexity arises when considering the solar system and the two near-Earth supernovae in conjunction with the large picture of the local Galactic neighborhood.  The solar system is currently inside of the Local Bubble, a region defined by low densities and high temperatures which is the result of numerous supernova explosions in the last 20 Myr \citep{frisch1981,smith2001,frisch2011,breitschwerdt2016,zucker2022}.  Although not inside the Local Bubble at the time of its formation, Earth crossed into the region around 5 Mya \citep{zucker2022}.  The timing places SN Plio (3 Mya) within the Local Bubble and therefore supernova remnant's expansion and properties should be considered in the context of a very low density ambient medium.  However, SN Mio (7 Mya) could have feasibly exploded outside the Local Bubble (thus expanding into a more general ISM medium) or, if it was inside, the \fe60-bearing dust grains would have had to cross the Local Bubble wall in order to reach the solar system.  Examining these potential differences and their effects in detail is beyond the scope of this work, however this interconnected picture is something to keep in mind in further studies.

\section{Conclusions}\label{sec:conc}

The distance to the 3 Mya supernova was last calculated using only two \fe60 signal measurements \citep{fry2015}.  With the large range of new data \citep{ludwig2016,fimiani2016,wallner2016,wallner2021}, it is the perfect time to update this distance.  We have also taken the opportunity to perform a parameter study on the astrophysical aspects of this problem and explore their effects on the supernova distance.  The main points are listed below.

\begin{itemize}

    \item We have evaluated the distance to SN Plio using all available \fe60 fluence data.  This allows us to examine the consistency among these measurements.  Comparison among results hinges on the adopted uptake or incorporation efficiency, which varies among sites and groups; more study here would be useful.  We find broad agreement among measurements, some of which are independent.  

    \item Fixing $\mej=3\times10^{-5}$ \msol\ and $\df = 10\%$, we find the distance to SN Plio (3 Mya) is $D \sim 20 - 140$ pc.  The distance to SN Mio (7 Mya) for the same astronomical parameters is $D \sim 110$ pc --- further variation is expected in this distance once more data have been analyzed.
    
    \item While the range quoted above for SN Plio covers the full potential range of the data, more realistically the distance to SN Plio is between $50 - 65$ pc.  This accounts for the measurements by \citet{wallner2016,wallner2021} and \citet{fitoussi2008}, falls inside the lunar range \citep{fimiani2016}, and is the approximate distance to Tuc-Hor, the stellar association most likely to host the CCSN \citep{fry2015,fry2020,hyde2018}. 
    
    \item \citet{wallner2021} measured both near-Earth supernova \fe60 signals in the same FeMn crust sample; by using a ratio of the fluences from these measurements, we can portion the excess \fe60 signal seen in the lunar regolith \citep{fimiani2016} into the contributions from the two supernovae.  We find that about 90\% of the lunar signal is from SN Plio and about 10\% is from SN Mio. 
    
    \item The sediment \fe60 detections from \citet{ludwig2016} are a valuable contribution to the field; unfortunately, a distance calculation reveals that the fluence quoted in their work produces an unrealistically far distance.  We must therefore suggest that their assumption of a 100\% uptake factor was incorrect or that some of the \fe60 in their samples was discarded. 
    
    \item The possible change in uptake factor for the \citet{knie2004} data accounts for the entirety of the $20 - 140$ pc range quoted for SN Plio --- efforts to constrain this uptake factor will be of great value to the field and help narrow down the Earth-based spread in the distance range.
    
    \item The astronomical parameters of ejected \fe60 mass and dust fraction have significant influence on the supernova distance and their possible values cover a wide range.  We can say that combinations of low \mej\ and low \df\ produce unrealistically close distances, while combinations of high \mej\ and high \df\ lead to unrealistically far distances.  For a supernova at about 50 pc, the combined parameter $\astropar\ \simeq 2\times10^{-6} \, \msol$. 
    Future observations that can shed new light on these parameters include CCSN dust measurements by JWST, and \fe60 gamma-ray line measurements by the upcoming COSI mission \citep{tomsick2022}.    
    
    \item The travel time parameter (how long the \fe60 is traveling from production to deposition on Earth) has a negligible effect on the supernova distance.
\end{itemize}

The plethora of work that has gone into analyzing the \fe60 signals over the last seven years has greatly increased our understanding of the two near-Earth supernovae; further efforts will help constrain the geophysical parameters in the distance calculations.  We are especially interested in results from sediment samples, as they are easiest to relate to the fluence of \fe60.  Furthermore, sediment samples of the 7 Mya SN will allow us to focus more closely on the astronomical differences between the two supernovae.  

The field of supernova dust dynamics is an active area of research and applies to far more than what we have summarized in this paper.  We look forward to advancements in the understanding of dust survival and destruction within remnants, as constraints on these numbers will help narrow our distance range.  Investigations into \fe60 production within supernovae and simulations of the explosion can also help tighten values for the \fe60 ejected mass.  Surveys, maps, and exploration of the Local Bubble allow further constraints on the distances to SN Plio and SN Mio.  Observations of dust in supernovae, such as with JWST, can probe grain production and evolution.  These independent measurements are invaluable, as they deal directly with the local neighborhood but are not correlated to the \fe60 signals detected on Earth.

Finally, and regardless of any and all possible limiting effects, enough \fe60 must 
travel from the supernovae to Earth to be detected by precision AMS measurements \textit{at least twice over} in the last 10 Myr.  
That such a signal has been observed twice in the (relatively) recent geologic past that the process of getting the \fe60 to Earth cannot overly  inhibiting, and indeed suggests that the interstellar spread of supernova radioisotope ejecta may be a robust process.

\acknowledgments
We are grateful to Shawn Bishop, Thomas Faestermann, Caroline Fitoussi, Jenny Feige, Gunther Korschinek, Peter Ludwig, and Toni Wallner for answering our questions about their data.
It is a pleasure to acknowledge useful discussions with Jesse Miller, Carla Fr\"{o}hlich, Sanjana Curtis, Zhenghai Liu, and Phil Coady.  The work of AFE and BDF was supported in part by the NSF under grand number AST-2108589, and benefitted from Grant No.~PHY-1430152 (JINA Center for the Evolution of the Elements).

\bibliography{references_3.bib}

\end{document}